\documentclass[referee,twocolumn, times, linenumbers, tighten, a4paper,fleqn,usenatbib,useAMS]{raa_twocolumn}           
\usepackage{graphicx,times}
\usepackage{natbib}
\usepackage{xcolor}
\usepackage{mathtools}
\usepackage{adjustbox}
\usepackage[version=4]{mhchem}
\usepackage{amssymb,amsmath}
\bibpunct{(}{)}{;}{a}{}{,}

\usepackage[colorlinks=true, allcolors=blue]{hyperref}

\begin{document}

\title{Observation of Complex Organic Molecules Containing Peptide-like Bonds
	Toward Hot Core G358.93--0.03 MM1}

 \volnopage{ {\bf 20XX} Vol.\ {\bf X} No. {\bf XX}, 000--000}
   \setcounter{page}{1}

   \author{Arijit Manna$^{*}$
   \inst{1}, Sabyasachi Pal\inst{1}}
%% Here is an example of three authors come from different institutes.
%% For single author or all the authors from an institute, use "\inst{}" only

   \institute{ $^{1}$Department of Physics and Astronomy, Midnapore City College, Paschim Medinipur, West Bengal, India 721129, {\it Email: amanna.astro@gmail.com}\\
   %$^{2}$S. N. Bose National Centre for Basic Sciences, Block-JD, Sector-III, Salt Lake City, Kolkata, India 700106\\
%% Please give the E-mail address of the author, to whom future correspondence and
%% offprint requests will be sent.
        %\and
          %   Yunnan Astronomical Observatory, National Astronomical Observatories, Chinese Academy of Sciences,
          %   Kunmin 650011, China\\
	%\and
%	  Center for Astrophysics, University of Science and Technology of China, Hefei 230026, China\\
%Key Laboratory for Research in Galaxies and Cosmology, The University of Science
%and Technology of China, Chinese Academy of Sciences, Hefei, Anhui, 230026, China\\
%\and 
%Polar Research Institute of China,
%Jinqiao Rd. 451, Shanghai, 200136, China\\
\vs \no
   {\small Received 2024 April 26; revised 2024 May 31; accepted 2024 June 1; published 2024 July 3}\\
   {\small DOI: 10.1088/1674-4527/ad539b}
}

\abstract{In star formation regions, the complex organic molecules (COMs) that contain peptide bonds (--NH--C(=O)--) play a major role in the metabolic process because --NH--C(=O)-- is connected to amino acids (R-CH{NH$_{2}$}--COOH). Over the past few decades, many COMs containing peptide-like bonds have been detected in hot molecular cores (HMCs), hot corinos, and cold molecular clouds, however, their prebiotic chemistry is poorly understood. We present the first detection of the rotational emission lines of formamide (NH$_{2}$CHO) and isocyanic acid (HNCO), which contain peptide-like bonds toward the chemically rich HMC G358.93--0.03 MM1, using high-resolution and high-sensitivity Atacama Large Millimeter/submillimeter Array bands 6 and 7. We estimate that the column densities of NH$_{2}$CHO and HNCO toward G358.93--0.03 MM1 are (2.80$\pm$0.29)$\times$10$^{15}$ cm$^{-2}$ and (1.80$\pm$0.42)$\times$10$^{16}$ cm$^{-2}$ with excitation temperatures of 165 $\pm$ 21 K and 170 $\pm$ 32 K, respectively. The fractional abundances of NH$_{2}$CHO and HNCO toward G358.93--0.03 MM1 are (9.03$\pm$1.44)$\times$10$^{-10}$ and (5.80$\pm$2.09)$\times$10$^{-9}$. We compare the estimated abundances of NH$_{2}$CHO and HNCO with the existing three-phase warm-up chemical model abundance values and notice that the observed and modeled abundances are very close. We conclude that NH$_{2}$CHO is produced by the reaction of NH$_{2}$ and H$_{2}$CO in the gas phase toward G358.93--0.03 MM1. Likewise, HNCO is produced on the surface of grains by the reaction of NH and CO toward G358.93--0.03 MM1. We also find that NH$_{2}$CHO and HNCO are chemically linked toward G358.93--0.03 MM1.
\keywords{ISM: individual objects (G358.93--0.03) -- ISM: abundances -- ISM: kinematics and dynamics -- stars: formation -- astrochemistry
}
}

   \authorrunning{Manna \& Pal}            %author_head in even pages
   \titlerunning{Peptide bond molecules in the G358.93--0.03 MM1}  % title_head in odd pages
   \maketitle

%________________________________________________ sections below
% 
\section{Introduction}
\label{sec:intro} 
In complex organic molecules (hereafter COMs), the peptide bond [--NH--C(=O)--] links two amino acids to produce proteins. So, those COMs with peptide bonds provide information regarding the origin of life in the universe. In the interstellar medium (ISM), formamide (NH$_{2}$CHO), cyanamide (NH$_{2}$CN), isocyanic acid (HNCO), and acetamide (CH$_{3}$CONH$_{2}$) are COMs containing peptide-like bonds. \ce{NH2CHO} has an amide-amide bond (--N--C(=O)--), which is crucial for the synthesis of proteins. NH$_{2}$CHO contains carbon (C), hydrogen (H), oxygen (O), and nitrogen (N), which are important for biological systems. NH$_{2}$CHO also acts as a prebiotic precursor of genetic and metabolic materials \citep{sal12}. The dipole moment of NH$_{2}$CHO is $\mu_{a}$ = 3.61 Debye and $\mu_{b}$ = 0.852 \citep{kur57}. The emission lines of NH$_{2}$CHO were first detected in high-mass star-formation regions Sgr B2 and Orion KL \citep{ru71}. The evidence of NH$_{2}$CHO was also found in several high- and low-mass star-formation regions \citep{bis07, ad13, ka13, lo15, man24}, extragalactic sources \citep{mu13}, comets \citep{bro00, biv14, go15}, and shocked regions \citep{yam12, men14}. 

In ISM, isocyanic acid (HNCO) is the simplest organic molecule, consisting of four biogenic elements, C, H, O, and N, all of which are present in living bodies. HNCO also acts as a precursor of several prebiotic COMs, including those molecules linked with astrochemical and astrobiological interest, such as amino acids, nucleobases, and sugars \citep{gor20, fed20}. Laboratory experiments suggested that CH$_{4}$ and HNCO can produce peptide-bonded molecules in the solid state \citep{lig18}. HNCO is a prolate asymmetric top molecule, and its rotational levels exhibit hyperfine splitting owing to the nuclear spin of nitrogen \citep{ni95, lap07}. The dipole moment of HNCO is $\mu_{a}$ = 1.60 Debye and $\mu_{b}$ = 1.35 Debye \citep{ku71}. The evidence of HNCO was first found towards Sgr B2 \citep{syn72, ch86, ku96}. Additionally, evidence of HNCO has also been found towards hot cores and hot corinos \citep{bl87, van95, mac96, bis08, gor20, can21}, galactic molecular clouds \citep{jak84, zin00}, and translucent clouds \citep{tur99}. {\color{blue}Although there were many chemical modelling works to investigate the evolution of \ce{NH2CHO} and HNCO, the formation pathways of both molecules remain unclear \citep{gor20}}.

The most chemically rich objects in the ISM are the hot molecular cores (hereafter HMCs) \citep{bis07, bel13, man22a, man22b, man23b}. HMCs represent the early phases of high-mass star-forming regions. The typical masses of HMCs are $\geq$100 \(\textup{M}_\odot\), which indicates that HMCs are reservoirs of several types of COMs, including precursors of NH$_{2}$CH$_{2}$COOH \citep{van98}. Our current idea regarding high-mass star formation regions is incomplete due to a lack of a sufficient number of observations of this kind of source. Recent observational studies indicate the evolution of high-mass star-forming regions as follows: infrared dark clouds (IRDCs) $\rightarrow$ HMCs $\rightarrow$ hyper/ultra-compact H II regions $\rightarrow$ H II regions, which are surrounded by ionized high-mass stars \citep{beu07}. The middle stages between the IRDCs and ultracompact HII regions are known as the HMCs. Owing to the lack of UV photons, the temperatures of the central protostars in the HMCs rise because of the increase in the temperatures of the dust and gaseous material. In ISM, HMCs have a high gas density ($n_{{H_{2}}}$$\geq$10$^{6}$ cm$^{-3}$), a warm temperature ($\geq$100 K), and a compact and small source size ($\leq$0.1 pc) \citep{van98, wi14}. Previous molecular line surveys at millimeter and sub-millimeter wavelengths show that several types of COMs like CH$_{3}$OH, CH$_{3}$OCHO, CH$_{3}$OCH$_{3}$, CH$_{3}$CN, and C$_{2}$H$_{5}$CN are frequently found in different HMCs because of the high temperature and evaporation of H$_{2}$O and organic ice mantles (\citet{her09}, and references therein).

\begin{figure*}
	\centering
	\includegraphics[width=1.0\textwidth]{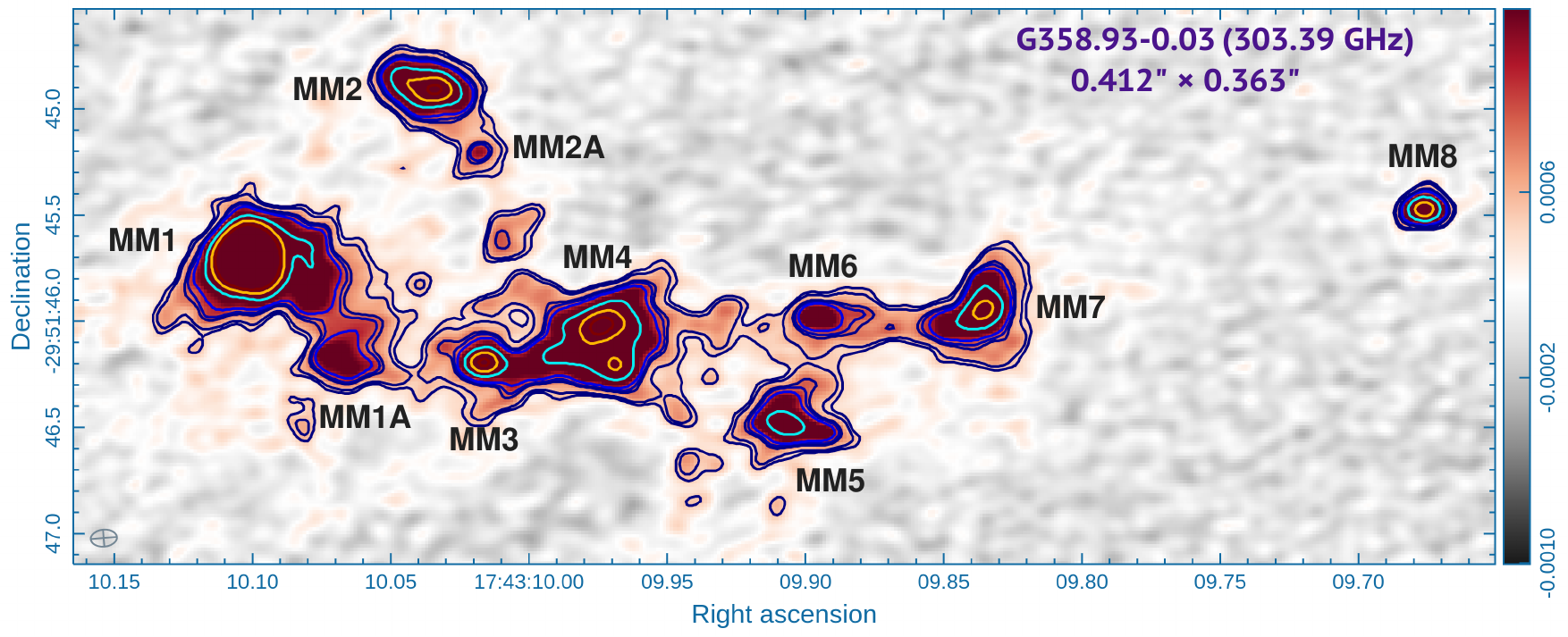}
	\caption{Millimeter wavelength continuum emission image of massive star-formation region G358.93--0.03 at a frequency of 303.39 GHz. The synthesized beam size of the image is 0.41$^{\prime\prime}$ $\times$ 0.36$^{\prime\prime}$. The contour levels start at 2.5$\sigma$ and increase by a factor of $\surd$2.}
	\label{fig:continuum}
\end{figure*}

G358.93--0.03 (RA: 17$^{h}$43$^{m}$10$^{s}$.02, Dec: -29$^{\circ}$51$^{\prime}$45.8$^{\prime\prime}$) is the massive star-forming region that was located 6.75$\,{}\,^{\,+0.37}_{\,-0.68}$\, kpc from Earth \citep{re14,bro19}. There are eight sub-millimeter continuum sources in G3589.93--0.03, which have been designated from G3589.93--0.03 MM1 to G3589.93--0.03 MM8 in decreasing right ascension (R.A) \citep{bro19}. The luminosity and mass of G358.93--0.03 are $\sim$7.7$\times$10$^{3}$ \textup{L}$_{\odot}$ and 167$\pm$12\textup{M}$_{\odot}$, respectively \citep{bro19}. Two of the eight sources, G358.93--0.03 MM1 (hereafter MM1) and G358.93--0.03 MM3 (hereafter MM3), have been confirmed as having line-rich hot molecular cores \citep{bro19, bay22}. Earlier, maser lines of \ce{CH3OH}, $^{13}$CH$_{3}$OH, HNCO, and HDO were detected towards MM1 using the ALMA, TMRT, and VLA radio telescopes \citep{bro19, chen20}. Except for the maser lines, \citet{bro19} also detected the emission lines of \ce{CH3CN} from MM1 and MM3, but they do not estimate the abundance of this molecule. The emission lines of NH$_{2}$CN, the simplest sugar-like molecule CH$_{2}$OHCHO, and the antifreeze molecule (CH$_{2}$OH)$_{2}$, and possible NH$_{2}$CH$_{2}$COOH precursor molecule CH$_{3}$NH$_{2}$ were also detected towards MM1 using the ALMA bands 6 and 7 \citep{man23a, man23b, man24b, man24a}. The detection of the above-mentioned molecules indicates that the MM1 is an ideal candidate for studying the emission lines of different types of complex biomolecules, including biologically relevant molecules. 

In this paper, we report the first detection of the emission lines of \ce{NH2CHO} and HNCO towards the hot molecular core MM1 with the ALMA. The gas temperature and column density of \ce{NH2CHO} and HNCO were estimated using the local thermodynamic equilibrium (LTE) model. We also address the probable formation pathways of \ce{NH2CHO} and HNCO towards MM1. The observations and data analysis are shown in section~\ref{obs}. The results of the emission line detection of NH$_2$CHO and HNCO are shown in Section~\ref{res}. The discussion and conclusions are presented in Sections ~\ref{dis} and \ref{conclu}.

\section{Observation and data reductions}
\label{obs}
We used openly available raw data of the massive star-forming region G358.93--0.03, observed with the Atacama Large Millimeter/Submillimeter Array (ALMA) 12-m arrays with bands 6 and 7 (which span the frequency ranges of 290.51--306.01 GHz and 225.44--242.06 GHz) (PI: Crystal Brogan). The band 6 observation was performed in four spectral windows with frequency ranges of 225.44--226.38 GHz, 229.48--229.72 GHz, 240.26--241.20 GHz, and 241.12--242.06 GHz. Similarly, the band 7 observation was performed in four spectral windows with frequency ranges of 290.51--292.39 GHz, 292.49--294.37 GHz, 302.62--304.49 GHz, and 304.14--306.01 GHz. The ALMA bands 6 and 7 observations were conducted on April 16, 2019 and November 11, 2019, with on-source integration times of 3265.92 sec and 756.0 sec. The star-formation region G358.93--0.03's phase centre was ($\alpha,\delta$)$_{\rm J2000}$ = 17:43:10.000, --29:51:46.000. To observe G358.93--0.03 in ALMA band 6, 45 antennas were set up with a minimum baseline of 15.1 m and a maximum baseline of 740.4 m. To observe G358.93--0.03 for ALMA band 7, 47 antennas were set up with a minimum baseline of 14 m and a maximum baseline of 2517 m. The bandpass and flux calibrators for bands 6 and 7 during the observations were J1924--2914 and J1550+0527, respectively, whereas the phase calibrator for bands 6 and 7 was J1744--3116.

We employed the Common Astronomy Software Application (CASA 5.4.1) with an ALMA data analysis pipeline for data reduction and imaging \citep{mc07}. We utilized the Perley-Butler 2017 flux calibrator model with the {\tt SETJY} task for flux calibration \citep{pal17}. After flagging the bad antenna data, we subsequently constructed the flux and bandpass calibration using pipeline tasks {HIFA\_BANDPASSFLAG} and {HIFA\_FLAGDATA}. We separated the target data with all available rest frequencies using the CASA task MSTRANSFORM after gain calibration. Utilizing the CASA task {TCLEAN} with line-free channels, we constructed a continuum emission image of G358.93--0.03. The continuum emission image of G358.93--0.03 at a frequency of 303.39 GHz is shown in Figure~\ref{fig:continuum}. Utilizing these data, \citet{man23b} recently investigated in detail the dust continuum emissions from G358.93--0.03 (see Figure 1 in \citet{man23b}). In addition to eight sources, \citet{man23b} also discovered two additional sources associated with MM1 and MM2, which are referred to as MM1A and MM2A. Thus, in this study, we did not address the continuum emission characteristics of G358.93--0.03. Next, we employ the UVCONTSUB task to remove the continuum emission from the UV plane of the calibrated data. After that, to produce spectral data cubes of G358.93--0.03, we used the CASA task {\tt TCLEAN} with the SPECMODE = CUBE parameter and Briggs weighting at a robust value of 0.5. Finally, we corrected the primary beam pattern in continuum images and spectral data cubes using the {\tt IMPBCOR} task.

\begin{figure*}
	\centering
	\includegraphics[width=1.0\textwidth]{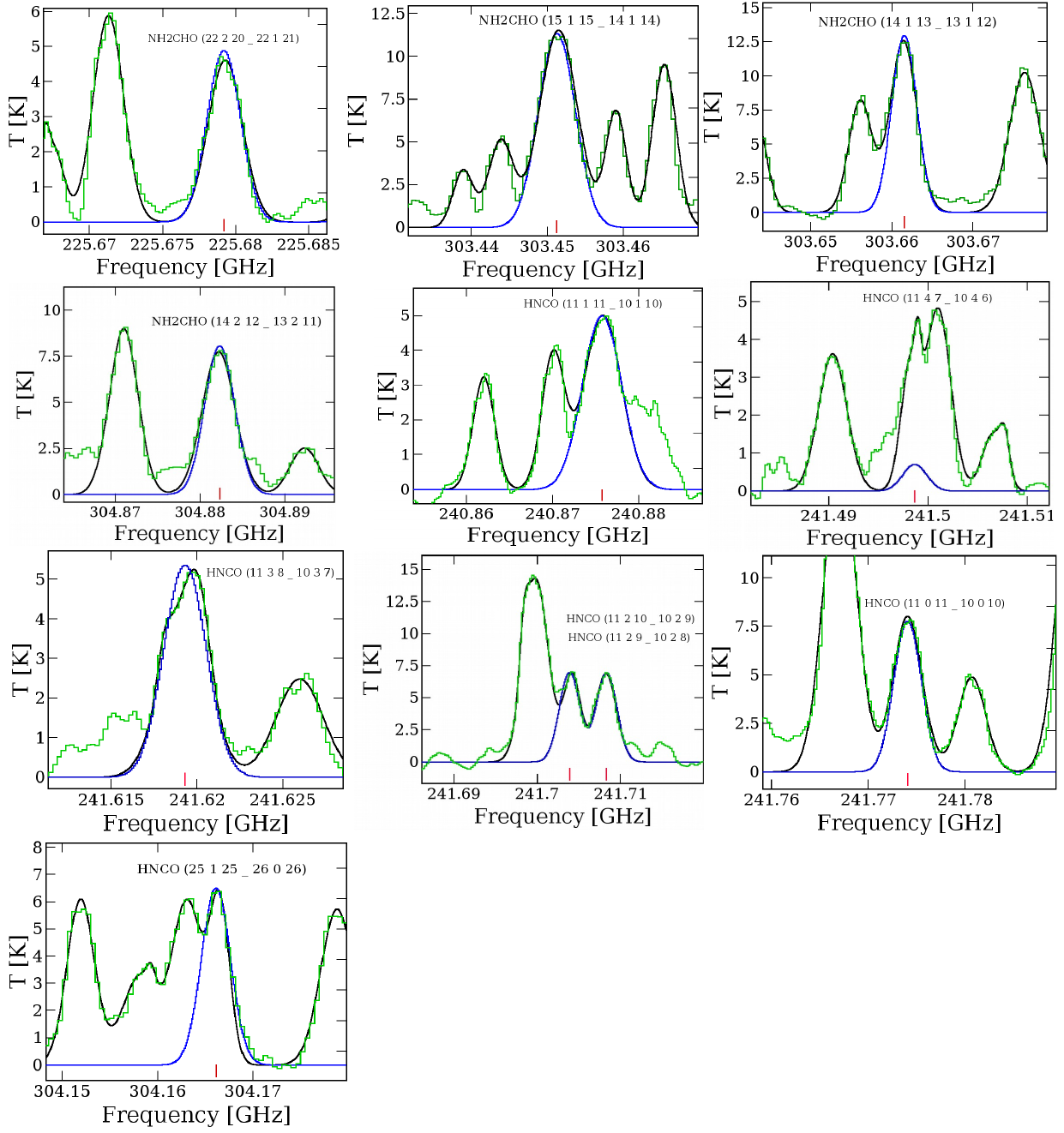}
	\caption{Emission lines of NH$_{2}$CHO and HNCO towards the MM1. The green lines represent the observed molecular spectra of MM1. The fitted blue spectra indicate the LTE-modelled spectra of \ce{NH2CHO} and HNCO. The fitted black spectra indicate the LTE-modelled spectra of other nearby molecules to understand the blended effect. The velocity of the spectra is --16.50 km s$^{-1}$.}
	\label{fig:line}
\end{figure*}

\section{Result}
\label{res}
\subsection{Molecular lines in G358.93--0.03}
In the spectral data cubes of G358.93--0.03, we observed molecular line emissions from hot cores MM1 and MM3. We did not find any molecular line emissions from other sources in G358.93--0.03. The synthesized beam sizes of the spectral data cubes of G358.93--0.03 at ALMA band 6 were 0.55$^{\prime\prime}\times$0.45$^{\prime\prime}$, 0.53$^{\prime\prime}\times$0.46$^{\prime\prime}$, 0.52$^{\prime\prime}\times$0.42$^{\prime\prime}$, and 0.51$^{\prime\prime}\times$0.42$^{\prime\prime}$. Similarly, the synthesized beam sizes of the spectral data cubes at ALMA band 7 were 0.42$^{\prime\prime}\times$0.36$^{\prime\prime}$, 0.42$^{\prime\prime}\times$0.37$^{\prime\prime}$, 0.41$^{\prime\prime}\times$0.36$^{\prime\prime}$, and 0.41$^{\prime\prime}\times$0.35$^{\prime\prime}$, respectively. The molecular spectra were extracted by drawing a 0.90$^{\prime\prime}$ diameter circular region over the MM1 and MM3, which is greater than the line emitting regions of both hot cores. The pointing centre of MM1 is $\alpha$ (J2000) = 17$^{h}$43$^{m}$10$^{s}$.101, $\delta$ (J2000) = --29$^\circ$51$^{\prime}$45$^{\prime\prime}$.693. Similarly, the pointing centre of MM3 is $\alpha$ (J2000) = 17$^{h}$43$^{m}$10$^{s}$.0144, $\delta$ (J2000) = --29$^\circ$51$^{\prime}$46$^{\prime\prime}$.193. Recently, \citet{man23b} used this band 7 data, and they claimed that MM1 is more chemically rich than MM3. In the band 7 molecular spectra of MM1, \citet{man23b} noticed an inverse P-cygni profile in the emission lines of \ce{CH3OH}. That indicates MM1 is undergoing an infall. \citet{man23b} did not see any inverse P-cygni profile in the emission lines of \ce{CH3OH} towards MM3. The velocities of the spectra ($V_{LSR}$) of MM1 and MM3 are --16.5 km s$^{-1}$ and --18.2 km s$^{-1}$, respectively, \citep{man23b, man24a}. From the chemically rich spectra, we focused only on studying the emission lines of COMs that contain peptide-like bonds, which are discussed in the following sections.

\begin{table*}
	\centering
	\scriptsize 
	\caption{Molecular line properties of NH$_{2}$CHO and HNCO towards the MM1.}
	\begin{adjustbox}{width=1.0\textwidth}
		\begin{tabular}{ccccccccccccccccccc}
			\hline 
			Molecule&Frequency &Transition&$E_{u}$ & $A_{ij}$ &g$_{up}$&$S\mu^{2}$&FWHM&$\rm{\int T_{mb}dV}$&Optical depth&Remarks\\
			&(GHz)&(${\rm J^{'}_{K_a^{'}K_c^{'}}}$--${\rm J^{''}_{K_a^{''}K_c^{''}}}$)&(K)&(s$^{-1}$)& &(Debye$^{2}$)&(km s$^{-1}$)&(K km s$^{-1}$)  &($\tau$)& \\
			\hline
			NH$_{2}$CHO&225.679&22(2,20)--22(1,21)&275.80&3.20$\times$10$^{-5}$&45&10.762&3.55$\pm$0.82&13.57$\pm$0.43 &1.37$\times$10$^{-1}$&Non blended\\

			&303.451&15(1,15)--14(1,14)&120.01&2.05$\times$10$^{-3}$&93&585.30 &3.58$\pm$0.29&44.61$\pm$0.96 &	2.04$\times$10$^{-1}$&Non blended\\
			
			&303.661&14(1,13)--13(1,12)&113.00&2.04$\times$10$^{-3}$ &87&545.30 &3.56$\pm$0.37&42.71$\pm$0.34 &1.99$\times$10$^{-1}$&Non blended\\
			
			&304.882&14(2,12)--13(2,11)&120.54&2.04$\times$10$^{-3}$ &87&538.17&3.55$\pm$0.44&27.44$\pm$0.52 &1.88$\times$10$^{-1}$&Non blended\\
			\hline
			HNCO&240.875&11(1,11)--10(1,10)&112.64&1.90$\times$10$^{-4}$&23&26.922&3.55$\pm$0.62&20.56$\pm$0.82 &1.98$\times$10$^{-1}$ &Non blended\\
			
			&~~241.498$^{*}$&11(4,7)-10(4,6)&720.30 &1.40$\times$10$^{-4}$&23&19.679&--&--&1.04$\times$10$^{-2}$&Blended with CHDCO\\
			
			&~~241.619$^{*}$&11(3,8)--10(3,7)&444.56&1.63$\times$10$^{-4}$&23&22.877&3.58$\pm$0.59&18.45$\pm$0.34   &8.55$\times$10$^{-2}$ &Non blended\\
			
			&~~241.703$^{*}$&11(2,10)--10(2,9)&239.89&1.81$\times$10$^{-4}$&23&25.357&3.49$\pm$0.21&15.28$\pm$0.22 &2.27$\times$10$^{-1}$&Non blended\\
			
			&~~241.708$^{*}$&11(2,9)--10(2,8)&239.89&1.81$\times$10$^{-4}$&23&25.356 &3.48$\pm$0.35&20.67$\pm$0.61 &2.37$\times$10$^{-1}$&Non blended\\
			
			&241.774&11(0,11)--10(0,10)&69.62&1.96$\times$10$^{-4}$&23&27.458&3.52$\pm$0.69&29.54$\pm$0.87 &5.69$\times$10$^{-1}$&Non blended\\ 
			
			&304.166&25(1,25)-26(0,26)&384.78&1.50$\times$10$^{-4}$&51&23.34&3.55$\pm$0.92&15.81$\pm$0.21 &1.13$\times$10$^{-1}$&Non blended\\
			
			\hline
		\end{tabular}	
	\end{adjustbox}
	\label{tab:MOLECULAR DATA}\\
	*--There are two transitions less than 100 kHz. We showed only the first transition.
	%		\end{minipage}[t]{\columnwidth}
\end{table*}

\subsection{Rotational emission lines NH$_{2}$CHO towards MM1}
We employed the local thermodynamic equilibrium (LTE) model with the Cologne Database for Molecular Spectroscopy (CDMS) and Jet Population Laboratory (JPL) molecular databases to find the rotational emission lines of NH$_{2}$CHO from the chemically rich molecular spectra of MM1 \citep{mu05, pic88}. For fitting the LTE spectral model over the observed emission lines of \ce{NH2CHO}, we employed the Markov Chain Monte Carlo (MCMC)\footnote{\url{https://cassis.irap.omp.eu/docs/CassisScriptingDoc/computation/mcmc_method.html\#mcmc-method}} algorithm in the CASSIS software package \citep{vas15}. The gas density of the inner region of MM1 is 2$\times$10$^{7}$ cm$^{-3}$, indicating that the LTE assumptions are true for this source \citep{ste21}. Through analysing the molecular spectra using the LTE model, we have detected four rotational emission lines of the complex N- and O-bearing molecule NH$_{2}$CHO towards MM1. The upper state energies of the detected emission lines of NH$_{2}$CHO varied between 113 K and 275.80 K. After detecting the emission lines of NH$_{2}$CHO, we obtained the molecular transitions, upper state energy ($E_u$) in K, Einstein coefficients ($A_{ij}$) in s$^{-1}$, line intensity ($S\mu^{2}$) in Debye$^{2}$, full-width half maximum (FWHM) in km s$^{-1}$, optical depth ($\tau$), and integrated intensities ($\rm{\int T_{mb}dV}$) in K km s$^{-1}$. The LTE fitting spectral lines of NH$_{2}$CHO are shown in Figure~\ref{fig:line} and the corresponding spectral parameters are listed in Table~\ref{tab:MOLECULAR DATA}. Except for NH$_{2}$CHO, we also fitted the rest of the 250 molecular transitions, which were taken from CDMS and JPL molecular databases, across the observed spectra of MM1 to better understand the blending effect. After spectral analysis, we noticed that all detected lines of NH$_{2}$CHO are non-blended, and all transition lines exhibit $\geq$4 $\sigma$ statistical significance. The best-fit column density of NH$_{2}$CHO is (2.80$\pm$0.29)$\times$10$^{15}$ cm$^{-2}$ with an excitation temperature of 165$\pm$21 K, which was estimated based on LTE modelling. The estimated excitation temperature of NH$_{2}$CHO suggests this molecule originates from the warm inner region of MM1 because the gas temperature of the warm inner region of the hot molecular cores is above 100 K \citep{van98, wi14}. The abundance of NH$_{2}$CHO towards MM1 with respect to molecular H$_{2}$ is (9.03$\pm$1.44)$\times$10$^{-10}$, where the column density of molecular H$_{2}$ towards MM1 is (3.10$\pm$0.2)$\times$10$^{24}$ cm$^{-2}$ \citep{man23b}.

\begin{figure*}
	\centering
	\includegraphics[width=1.0\textwidth]{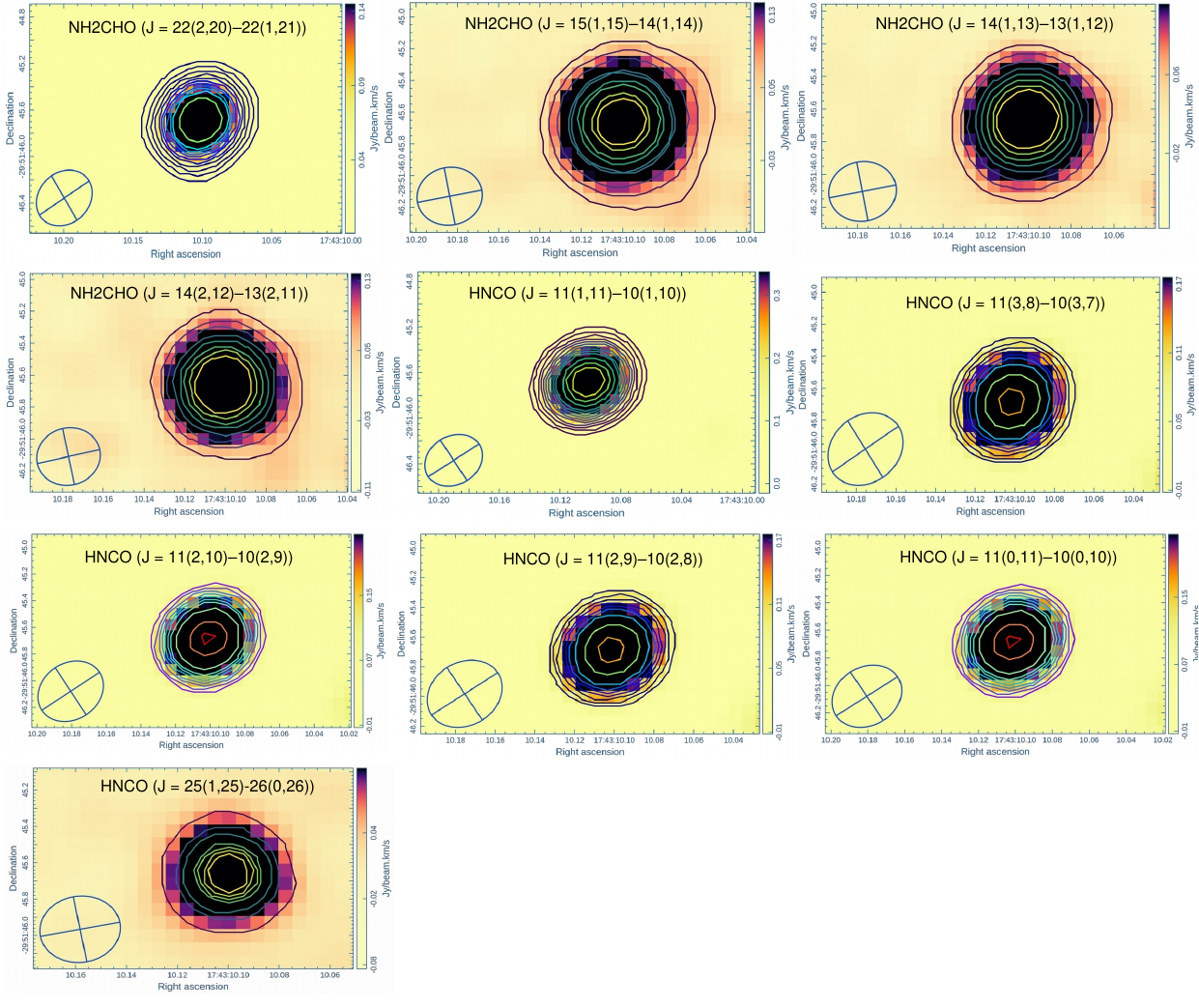}
	\caption{Emission maps of detected non-blended transitions of NH$_{2}$CHO and HNCO towards the MM1. The contours are the 988 $\mu$m continuum emission map of G358.93--0.03. The contour levels start at 3$\sigma$.}
	\label{fig:emissionmap}
\end{figure*}

\subsection{Rotational emission lines of HNCO towards MM1}
Using the LTE-modelled spectra, we detected seven rotational emission lines of HNCO towards MM1 using the ALMA bands 6 and 7. The upper-state energies ($E_{u}$) of the identified rotational emission lines of HNCO vary between 69.62 K and 720.30 K. After spectral analysis, we noticed that the emission line of HNCO at a frequency of 241.498 GHz (J = 11(4,7)--10(4,6)) is blended with CHDCO. All other detected transitions are non-blended. The non-blended emission lines of HNCO exhibited $\geq$4.5 $\sigma$ statistical significance. The resultant LTE-fitted spectral line and spectra line parameters of HNCO are shown in Figure~\ref{fig:line} and Table~\ref{tab:MOLECULAR DATA}. The best-fit column density of HNCO is (1.80$\pm$0.42)$\times$10$^{16}$ cm$^{-2}$ with an excitation temperature of 170$\pm$32 K, which was estimated based on LTE modelling. The abundance of HNCO with respect to molecular H$_{2}$ towards the MM1 is (5.80$\pm$2.09)$\times$10$^{-9}$, where the column density of molecular H$_{2}$ towards the MM1 is (3.10$\pm$0.2)$\times$10$^{24}$ cm$^{-2}$ \citep{man23b}.

\subsection{Searching the emission lines of NH$_{2}$CHO and HNCO towards MM3}
After the detection of the rotational emission lines of NH$_{2}$CHO and HNCO towards MM1, we used the LTE-modelled spectra to search for these molecules towards MM3. However, we were unable to find the emission lines of those molecules towards MM3. As per the LTE analysis, the upper limit column densities of NH$_{2}$CHO and HNCO towards MM3 are $\leq$(2.05$\pm$0.85)$\times$10$^{13}$ cm$^{-2}$ and $\leq$(8.12$\pm$1.26)$\times$10$^{14}$ cm$^{-2}$, respectively. The upper limit of the abundances of NH$_{2}$CHO and HNCO towards MM3 are $\leq$(5.84$\pm$1.26)$\times$10$^{-11}$ and $\leq$(2.31$\pm$1.27)$\times$10$^{-9}$, where the column density of molecular H$_{2}$ towards the MM3 is (3.51$\pm$0.7)$\times$10$^{23}$ cm$^{-2}$ \citep{man23b}.

\subsection{Spatial distribution of NH$_{2}$CHO and HNCO towards MM1}
\label{sec:fitting}
After detection of the emission lines of NH$_{2}$CHO and HNCO towards MM1, we produced the integrated emission maps of the non-blended emission lines of those molecules by integrating the spectral data cubes using the channel ranges with the help of CASA task {\tt IMMOMENTS}. The emission maps of NH$_{2}$CHO and HNCO towards MM1 are presented in Figure~\ref{fig:emissionmap}. The emission maps of NH$_{2}$CHO and HNCO are overlaid with a 988 $\mu$m continuum emission image of G358.93--0.03. The dust continuum image of G358.93--0.03 is taken from \citet{man23b}. From the integrated emission maps, we observed that NH$_{2}$CHO and HNCO arose from the warm-inner regions of MM1. Since the excitation temperature of \ce{NH2CHO} and HNCO towards MM1 is 165$\pm$21 K and 170$\pm$32 K, which indicates both molecules may exist in the same region of MM1. So, there is a high chance of a chemical link between \ce{NH2CHO} and HNCO towards MM1. We estimate the size of the emitting regions of NH$_{2}$CHO and HNCO by fitting the 2D Gaussians over the emission maps of NH$_{2}$CHO and HNCO using the {\tt IMFIT} task. The sizes of the emitting regions of NH$_{2}$CHO and HNCO corresponding to the non-blended transition lines are shown in Table~\ref{tab:emittingregion}. We observed that the emitting regions of NH$_{2}$CHO and HNCO are similar or marginally larger than the beam sizes of the emission maps. This suggests that the emission maps of NH$_{2}$CHO and HNCO are not spatially resolved towards MM1. Therefore, we could not determine the chemical morphologies of NH$_{2}$CHO and HNCO from the integrated emission maps. Higher spatial and angular resolution observations are needed to learn the spatial distribution of NH$_{2}$CHO and HNCO towards MM1. 

\begin{table}{}
	%\scriptsize
	%	\begin{minipage}[t]{\columnwidth}
	\centering
	\caption{Emitting regions of NH$_{2}$CHO and HNCO towards the MM1.}
	\begin{adjustbox}{width=0.65\textwidth}
		\begin{tabular}{cccccccccccc}
			\hline
			Molecule&Frequency&Transition& E$_{up}$&Emitting region\\
			
			&GHz&(${\rm J^{'}_{K_a^{'}K_c^{'}}}$--${\rm J^{''}_{K_a^{''}K_c^{''}}}$)&(K)&($^{\prime\prime}$)\\
			\hline
			NH$_{2}$CHO& 225.679&22(2,20)--22(1,21)&275.80&0.42\\
			&303.451&15(1,15)--14(1,14)&120.01&0.43\\
			&303.661&14(1,13)--13(1,12)&113.00&0.42\\
			&304.882&14(2,12)--13(2,11)&120.54&0.43\\
			\hline
			HNCO&240.875&11(1,11)--10(1,10)&112.64&0.42\\
			&~~241.619$^{*}$&11(3,8)--10(3,7)&444.56&0.42\\
			&~~241.703$^{*}$&11(2,10)--10(2,9)&239.89&0.43\\
			&~~241.708$^{*}$&11(2,9)--10(2,8)&239.89&0.43\\
			&241.774      &11(0,11)--10(0,10)&69.62&0.43\\
			&304.166      &25(1,25)--26(0,26)&384.78&0.44\\
			\hline
		\end{tabular}	
	\end{adjustbox}
	
	*--There are two transitions within less than 100 kHz. We show only the first transition.\\
	
	\label{tab:emittingregion}
	%	\end{minipage}[t]{\columnwidth}
\end{table}	

\subsection{Searching of other peptide bonds molecules towards MM1}
After the detection of the rotational emission lines of NH$_{2}$CHO and HNCO towards MM1, we also search the rotational emission lines of other COMs that have peptide bonds, such as HOCH$_{2}$C(O)NH$_{2}$, CH$_{3}$NCO, CH$_{3}$NHCHO, NH$_{2}$C(O)CN, CH$_{3}$CH$_{2}$NCO, CH$_{3}$C(O)NH$_{2}$, and NH$_{2}$C(O)NH$_{2}$ using the LTE modelled spectra. During the spectral analysis, we observed that all the above-mentioned molecules are blended with other molecular transitions. So, there is a very low chance of identifying these molecules. The upper-limit column densities of peptide bond molecules HOCH$_{2}$C(O)NH$_{2}$, CH$_{3}$NCO, CH$_{3}$NHCHO, NH$_{2}$C(O)CN, CH$_{3}$CH$_{2}$NCO, CH$_{3}$C(O)NH$_{2}$, and NH$_{2}$C(O)NH$_{2}$ are $\leq$3.56$\times$10$^{14}$ cm$^{-2}$, $\leq$8.26$\times$10$^{14}$ cm$^{-2}$, $\leq$5.20$\times$10$^{13}$ cm$^{-2}$, $\leq$2.96$\times$10$^{13}$ cm$^{-2}$, $\leq$9.62$\times$10$^{13}$ cm$^{-2}$, $\leq$1.16$\times$10$^{14}$ cm$^{-2}$, and $\leq$9.28$\times$10$^{12}$ cm$^{-2}$, respectively.

\section{Discussion}
\label{dis}
\subsection{Comparison of the abundances of \ce{NH2CHO} and HNCO towards MM1 and other sources}
To understand the distribution of NH$_{2}$CHO and HNCO in the ISM, we compare the fractional abundance of those molecules towards MM1 and other hot molecular cores such as G10.47+0.03, G31.41+0.31, Orion KL, Sgr B2, and hot corino object IRAS 16293--2422 B. This comparison is presented in Table~\ref{tab:comparision}. After comparison, we noticed that the fractional abundance of NH$_{2}$CHO towards MM1 is nearly similar to Sgr B2 and IRAS 16293--2422 B. The fractional abundance of NH$_{2}$CHO towards MM1 is one order of magnitude lower than the G31.41+0.31, G10.47+0.03, and Orion KL. Similarly, the fractional abundance of HNCO towards MM1 is similar to that of Orion KL and Sgr B2. The fractional abundance of HNCO is one order of magnitude lower than that of G10.47+0.03 and G31.41+0.31. Furthermore, we notice that the fractional abundance of HNCO towards MM1 is two orders of magnitude larger than that of IRAS 16293--2422 B. Based on that comparison, we identify that G31.41+0.31 and G10.47+0.03 are the most abundant NH$_{2}$CHO and HNCO sources in the ISM. The low abundances of NH$_{2}$CHO and HNCO towards MM1 indicate that the detection probabilities of other peptide bond COMs are very low in these sources. 

\begin{table*}
	\centering
	\scriptsize 
	\caption{Comparision of the abundances of NH$_{2}$CHO and HNCO towards the MM1 and other sources.}
	\begin{adjustbox}{width=1.0\textwidth}
		\begin{tabular}{ccccccccccccccccc}
			\hline 
			Molecule&MM1$^{a}$&G10.47+0.03$^{b}$&G31.41+0.31$^{c}$&Orion KL$^{d}$&Sgr B2$^{e}$&IRAS 16293--2422 B$^{f}$\\
			\hline
			NH$_{2}$CHO&(9.03$\pm$1.44)$\times$10$^{-10}$&2.87$\times$10$^{-9}$&(5.4$\pm$1.5)$\times$10$^{-9}$&5.80$\times$10$^{-9}$& 1.3$\times$10$^{-10}$&1.05$\times$10$^{-10}$ \\
\hline			
			HNCO       &(5.80$\pm$2.09)$\times$10$^{-9}$ &1.02$\times$10$^{-8}$&(1.1$\pm$0.3)$\times$10$^{-8}$&4.08$\times$10$^{-9}$&7.20$\times$10$^{-9}$& (1.8$\pm$0.4)$\times$10$^{-11}$\\

			\hline
		\end{tabular}	
	\end{adjustbox}
	\label{tab:comparision}\\
	Notes: a: Our work, b: \citet{gor20}, c: \citet{col21}, d: \citet{ka13}, \citet{ter10}, e: \citet{hal11}, \citet{ad10}, f: \citet{ka13}, \citet{he19} \\
	%		\end{minipage}[t]{\columnwidth}
\end{table*}

\subsection{Possible formation mechanism of NH$_{2}$CHO and HNCO towards MM1}
The formation mechanism(s) of NH$_{2}$CHO in ISM is still debatable in the astrochemistry community. We observed that only a few reactions have been proposed to understand the formation pathways of NH$_{2}$CHO in both the grain surface and the gas phase. Previously, \citet{qu07} claimed that the NH$_{2}$CHO created via the ion-molecule reaction and subsequent electron recombination reaction between NH$_{4}$$^{+}$ and H$_{2}$CO:\\\\
NH$_{4}$$^{+}$ + H$_{2}$CO$\longrightarrow$ NH$_{3}$CHO$^{+}$ + H$_{2}$~~~~~~~~~~~~~(1)\\\\
NH$_{3}$CHO$^{+}$ + e$^{-}$$\longrightarrow$NH$_{2}$CHO + H~~~~~~~~~~~~~~~(2)\\\\
Subsequently, \citet{gar08} claimed that the rates of reactions 1 and 2 are unknown. In particular, \citet{gar08} proposed the following radical-neutral reaction in the gas phase:\\\\
H$_{2}$CO + NH$_{2}$$\longrightarrow$ NH$_{2}$CHO + H~~~~~~~~~~~~~~~~~~(3)\\\\
However, \citet{red14} showed that the net activation barrier energy of this reaction is above 1000 K. Earlier, \citet{gar13} and \citet{gor20} showed that reaction 3 is responsible for the formation of \ce{NH2CHO} towards hot molecular cores and hot corions. {\color{blue}Previously \citet{cou16} showed that \ce{NH2CHO} was formed via the reaction between \ce{H2CO} and \ce{NH2} (reaction 3) in the gas phase towards the hot corino IRAS 16293--2422 B. Similarly, reaction 3 is also responsible for the formation of \ce{NH2CHO} in the shock region L1157--B1, hot cores Sgr B2 (N), G10.47+0.03, Orion KL, G31.41+0.31, and the high-mass protostar IRAS 18089--1732 \citep{hal11, ka13, cod17, gor20, col21, man24}}.

Previous studies have proposed that the emission lines of HNCO originate from the high-density warm-inner regions of hot cores and hot corinos. Therefore, HNCO may play an important role as a tracer for dense gases in hot molecular cores \citep{jak84}. Previously, \citet{ig77} claimed that the HNCO produced in the gas phase via the ion-neutral reaction towards Sgr B2:\\\\
NCO$^{+}$ + H$_{2}$ $\longrightarrow$ HNCO$^{+}$ + H~~~~~~~~~~~~~~~~~~(4)\\\\
HNCO$^{+}$ + H$_{2}$ $\longrightarrow$ HNCOH$^{+}$ + H~~~~~~~~~~~~(5)\\\\
HNCOH$^{+}$ + e$^{-}$ $\longrightarrow$ HNCO + H~~~~~~~~~~~~~~(6)\\\\
Previously, \cite{tur00} suggested another neutral-neutral reaction to create HNCO:\\\\
CN + O$_{2}$ $\longrightarrow$ NCO + O~~~~~~~~~~~~~~~~~~~~~~~~~~~(7)\\\\
NCO + H$_{2}$ $\longrightarrow$ HNCO + H~~~~~~~~~~~~~~~~~~~~~(8)\\\\
Reaction 8 has an activation barrier. Previously, \citet{gar08} showed that the HNCO also formed on the grain surface via the thermal reaction:\\\\
NH + CO $\longrightarrow$ HNCO~~~~~~~~~~~~~~~~~~~~~~~~~~~~~(9)\\\\
\citet{gar08} claimed that reaction 9 has no activation barrier and that the molecule is released in the gas phase via desorption. Earlier, \citet{gar13} and \citet{gor20} show that reaction 9 is the most efficient reaction for the formation of HNCO towards the hot molecular cores.

Previously, \citet{gor20} computed a three-phase (gas + grain + ice mantle) warm-up chemical model to understand the possible formation mechanisms of NH$_{2}$CHO and HNCO. During chemical modelling, they assumed the free-fall collapse of a cloud (Phase I), which was followed by the warm-up phase. In phase I, the density of the gas rapidly increased from 10$^{3}$ cm$^{-3}$ to 1$\times$10$^{7}$ cm$^{-3}$ and the dust temperature was constant at 10 K. In phase II (the warm-up stage), the dust temperature was increased from 10 K to 400 K, and the density of the gas was constant at 1$\times$10$^{7}$ cm$^{-3}$. At this stage, the gas and dust in the hot core were well coupled. During chemical modelling, \citet{gor20} used the radical-neutral reaction between H$_{2}$CO and NH$_{2}$ (reaction 3) and the reaction between NH and CO (reaction 9) for the formation of NH$_{2}$CHO and HNCO in the gas phase and grain surface, respectively. Previously, \citet{gar13} also showed that reactions 3 and 9 are the most efficient for the formation of NH$_{2}$CHO and HNCO in the gas phase and grain surface towards hot molecular cores. In the warm-up stage, \citet{gor20} estimated the modelled abundances of NH$_{2}$CHO and HNCO towards hot molecular cores to be 8.27$\times$10$^{-10}$ and 4.8$\times$10$^{-9}$, respectively.

To understand the possible formation mechanism of NH$_{2}$CHO and HNCO towards MM1, we compared our derived abundance of NH$_{2}$CHO and HNCO with the modelled abundance of \citet{gor20}. This comparison is reasonable because the gas density and temperature of this source are 2$\times$10$^{7}$ cm$^{-3}$ and 150 K, respectively \citep{chen20, ste21}. Therefore, the three-phase warm-up chemical model of \citet{gor20} is efficient for understanding the formation pathways of NH$_{2}$CHO and HNCO towards the MM1. \citet{gor20} derived the modelled abundance of NH$_{2}$CHO and HNCO to be 8.27$\times$10$^{-10}$ and 4.8$\times$10$^{-9}$, respectively. The observed fractional abundances of NH$_{2}$CHO and HNCO towards MM1 are (9.03$\pm$1.44)$\times$10$^{-10}$ and (5.80$\pm$2.09)$\times$10$^{-9}$, which are nearly similar to the modelled abundance results in \citet{gor20}. This result indicates that NH$_{2}$CHO is produced by the reaction of NH$_{2}$ and H$_{2}$CO in the gas phase towards MM1. Similarly, HNCO is created by the reaction of NH and CO on the surface of the grains of the MM1.

\subsection{Chemical link between NH$_{2}$CHO and HNCO}
Several studies have shown that NH$_{2}$CHO and HNCO are chemically connected. Earlier, \citet{gar08}, \citet{hau19}, and \citet{gor20} claimed that the subsequently hydrogenation of HNCO creates NH$_{2}$CHO in the grain surface:\\\\
HNCO + H $\longrightarrow$ NH$_{2}$CO~~~~~~~~~~~~~~~~~~(10)\\\\
NH$_{2}$CO + H $\longrightarrow$ NH$_{2}$CHO~~~~~~~~~~~~~~(11)\\\\
In another way, \citet{lo15} demonstrated the molecular correlation between the abundance of HNCO and NH$_{2}$CHO towards the high-mass star-formation regions. From the molecular correlation plot, \citet{lo15} found a positive correlation between the abundances of \ce{NH2CHO} and HNCO, and they found a correlation equation $X$(NH$_{2}$CHO) = 0.04$X$(HNCO)$^{0.93}$ (see Figures 2 and 3 in \citet{lo15}). According to the molecular correlation equation of \citet{lo15}, we found that the abundance of NH$_{2}$CHO is (8.75$\pm$1.98)$\times$10$^{-10}$, which is very similar to our observed abundance of NH$_{2}$CHO towards MM1. This result indicates that HNCO and NH$_{2}$CHO are chemically connected towards MM1. 

\section{Conclusion}
\label{conclu}
In this article, we present the identification of the emission lines of HNCO and NH$_{2}$CHO towards hot core MM1 using the ALMA bands 6 and 7. The conclusions of this study are summarised below: \\\\
1. We successfully detected four and six non-blended rotational transition lines of peptide bond molecules NH$_{2}$CHO and HNCO towards the hot core MM1. \\\\
2. The estimated column densities of NH$_{2}$CHO and HNCO towards the MM1 are (2.80$\pm$0.29)$\times$10$^{15}$ cm$^{-2}$ and (1.80$\pm$0.42)$\times$10$^{16}$ cm$^{-2}$ with excitation temperatures of 165$\pm$21 K and 170$\pm$32 K, respectively. The fractional abundances of NH$_{2}$CHO and HNCO towards the MM1 are (9.03$\pm$1.44)$\times$10$^{-10}$ and (5.80$\pm$2.09)$\times$10$^{-9}$.\\\\
3. We compared the estimated abundances of NH$_{2}$CHO and HNCO with the existing three-phase warm-up chemical model abundances and noticed that the observed and modelled abundances are very close.\\\\
4. We claim that NH$_{2}$CHO is produced by the reaction of NH$_{2}$ and H$_{2}$CO in the gas phase towards MM1. Similarly, HNCO is created by the reaction of NH and CO on the grain surface of MM1. \\\\
5. We also discuss the chemical links between \ce{NH2CHO} and HNCO. According to the correlation equation of \citet{lo15}, both \ce{NH2CHO} and HNCO are chemically linked towards the MM1. The presence of \ce{NH2CHO} and HNCO indicates that other O- and N-bearing molecules exist in this source, which we discuss in follow-up studies.

\section*{acknowledgments} We thank the anonymous referee for the helpful comments that improved the manuscript.
 A.M. acknowledges the SVMCM for financial support for this research. This paper makes use of the following ALMA data: ADS /JAO.ALMA\#2019.1.00768.S and ADS /JAO.ALMA\#2018.A.00031.T. ALMA is a partnership of ESO (representing its member states), NSF (USA), and NINS (Japan), together with NRC (Canada), MOST and ASIAA (Taiwan), and KASI (Republic of Korea), in co-operation with the Republic of Chile. The joint ALMA Observatory is operated by ESO, AUI/NRAO, and NAOJ.

\section*{Conflicts of interest}
The authors declare no conflict of interest.

\bibliographystyle{aasjournal}
%\bibliography{./literature.bib,added.bib} % if your bibtex file is called example.bib

\end{document}